\RequirePackage{fix-cm}
\documentclass[final,3p,onecolumn,authoryear]{elsarticle}
\usepackage[table,dvipsnames]{xcolor}
\usepackage{graphicx}
\usepackage{amsmath}
\usepackage[unicode,pdfencoding=auto,hidelinks,pdfhighlight=/P]{hyperref}
\usepackage{natbib}

\hypersetup{linkcolor=Plum,urlcolor=NavyBlue,citecolor=OliveGreen,colorlinks=true}

\newcommand{\Gg}[2]{\ensuremath{#1^\circ\,\text{#2}}}
\newcommand{\E}[1]{\Gg{#1}{E}}

\newcommand{\N}[1]{\Gg{#1}{N}}
\newcommand{\EE}[2]{\Gg{#1}{E}\,\text{--}\,\Gg{#2}{E}}
\newcommand{\NN}[2]{\Gg{#1}{N}\,\text{--}\,\Gg{#2}{N}}
\newcommand{\mdash}{\,--\,}

\journal{Ocean Dynamics}
\begin{document}
\hypersetup{linkcolor=Plum,urlcolor=NavyBlue,citecolor=OliveGreen,colorlinks=true}

\begin{frontmatter}
\title{Mesoscale circulation along the Sakhalin Island eastern coast}

\author{S.V. Prants}
\ead{prants@poi.dvo.ru}
\ead[url]{dynalab.poi.dvo.ru}

\author{A.G. Andreev}
\author{M.V. Budyansky}
\author{M.Yu. Uleysky}

\address{Pacific Oceanological Institute of the Russian Academy of Sciences,\\
Laboratory of Nonlinear Dynamical Systems,\\
43 Baltiiskaya st., 690041 Vladivostok, Russia\\
URL: \url{http://dynalab.poi.dvo.ru}}

\begin{abstract}
The seasonal and interannual variability of mesoscale circulation along the eastern coast of the Sakhalin 
Island in the Okhotsk Sea is investigated using AVISO velocity field and oceanographic data for the period 
from 1993 to 2016. It is found that mesoscale cyclones with the horizontal dimension of about 100~km occur there 
predominantly during summer, whereas anticyclones occur predominantly during fall and winter. The cyclones are 
generated due to the coastal upwelling forced by northward winds and the positive wind stress curl along the 
Sakhalin coast. The anticyclones are formed due to an inflow of low-salinity Amur-River waters from the Sakhalin 
Gulf intensified by southward winds and the negative wind stress curl in the cold season. The mesoscale cyclones 
support the high biological productivity at the eastern Sakhalin shelf in July\mdash August.
\end{abstract}

\begin{keyword}
Okhotsk Sea \sep East Sakhalin Current \sep mesoscale circulation
cells and their seasonality
\end{keyword}
\end{frontmatter}

\section{Introduction}\label{intro}
\begin{figure*}[!htb]
\begin{center}
\includegraphics[width=0.75\textwidth,clip]{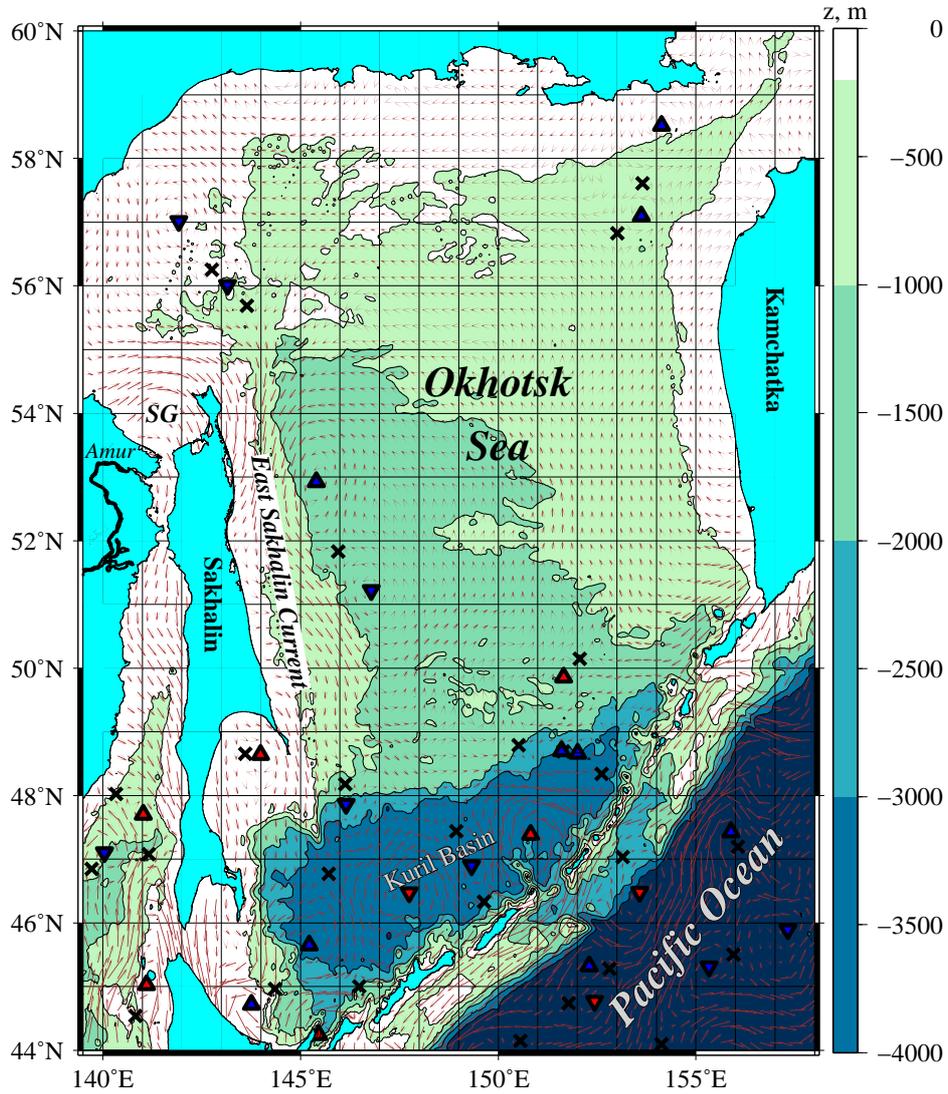}
\end{center}
\caption{The bathymetry of the Okhotsk Sea with the altimetric AVISO velocity field imposed (arrows) 
averaged for the period from 1993 to 2016. The elliptic and hyperbolic stagnation points with zero 
mean velocity are indicated by triangles and crosses, respectively. SG is for the Sakhalin Gulf.}
\label{fig1}
\end{figure*}

The Okhotsk Sea (OS) is one of the marginal seas in the North Pacific. It is bounded by the Kamchatka 
Peninsula, Siberia, Sakhalin Island (SI), Hokkaido and the Kuril Islands. The OS is connected to the 
subarctic North Pacific through the Kuril Islands chain. About 50--70\% of the OS area is covered with 
ice in winter. The distribution of dynamic topography in the OS indicates a general cyclonic circulation 
in the north and an anticyclonic circulation in the deep Kuril Basin in the south 
\citep{Moroshkin1966, Ohshima2004}. The East Sakhalin Current (ESC) is the western boundary current of 
the OS cyclonic gyre (Fig.~\ref{fig1}). The ESC transports southward along the eastern SI shelf-break 
low salinity surface water affected by the Amur-River discharge.

The intensity and direction of the ESC change seasonally. The seasonal maps of the geostrophical 
currents (relative to 500~dbar with the data collected between 1948 and 1994) demonstrate a strong
southward flux along the eastern SI shelf and slope in late fall and a weak northward flux along
200~m isobath in summer \citep{Pishchalnik1999}. The northward flow in the surface layer along 
the northeastern Sakhalin shelf (\N{52.5}) during August and first decade of September and the strong 
southward flow during the second and third decades of September 1997 and 1998 have been observed 
by \citet{Kochergin1999}. They indicated a positive correlation between meridional wind and 
meridional velocity (mooring data). \citet{Ohshima2004} have demonstrated by using the wind stress 
curl and mooring data that the computed Sverdrup transport and the observed ESC transport (\N{53}, July 
1998\mdash January 2000, depth ${\sim}100$~m) exhibit large seasonal variations with maximum in 
winter and minimum in summer. They assumed that the main part (the shelf-slope core) of the ESC 
can be regarded as a western boundary current of the wind-driven cyclonic gyre. The lack of the 
observed northward transport of the ESC across \N{53} in summer of 1999 during the period of the 
negative (anticyclonic) wind stress curl was explained by importance of the annually mean wind 
stress curl for the southward flow of the ESC in summer. \citet{Ebuchi2006} has studied the seasonal 
and interannual variations in the ESC and its relation to wind stress and wind stress curl fields 
in the OS using ten-year (1992--2002) records of the sea level anomaly observed by the TOPEX/POSEIDON 
altimeter. He concluded that the southward flow of the ESC is strong in winter and almost disappears 
in summer and that more observations are required to clarify the relationship between the interannual 
variations in the ESC and those of the wind fields over the OS.

Using the CTD data collected in summer 1994, \citet{Verkhunov1997} revealed the mesoscale cyclonic 
circulation off the northeastern SI and the northward transport of the ESC along the slope and the 
southward transport along the shelf. The existence of the northeastward and northwestward currents 
along the East Sakhalin slope (\N{51.5}) in summer 2009 and 2010 has been shown by 
\citet{Kusailo2013} using the mooring data.

The northeastern SI shelf is known as a region with remarkably high primary production of
1.5--2~g~C~m$^{-2}$~day$^{-1}$ in the post-spring bloom period (July\mdash August) \citep{Sorokin_2002}.
The physical processes at the northeastern SI shelf are of a special interest in view of the high 
productivity of benthos communities eaten here by grey whales during the annual summer-autumn fattening 
\citep{Meier2007}. The benthos fauna grow is due to a detritus flux provided by the phytoplankton bloom. 
The mesoscale anticyclones and cyclones, observed off the eastern SI, could have a profound effect on the 
physical and biological environments and can impact the marine ecosystem at the SI shelf from plankton 
distribution to higher trophic levels such as feeding and growth of eggs and larvae and benthos.

In this study we show that the mesoscale cyclones with the horizontal dimension of 100~km occur in 
the ESC region predominantly during summer, whereas anticyclones are generated predominantly during 
fall and winter (October\mdash December). The mesoscale cyclones generation is related to the coastal 
upwelling forcing by northward winds and positive wind stress curl along the SI coast. The anticyclones 
formation is related to inflow of low salinity waters from the Sakhalin Gulf driven by southward winds 
and negative wind stress curl along the SI coast.

\section{Data and methods}\label{data_methods}
We used the geostrophic daily velocities for the period from January~2, 1993 to March~17, 2016 
obtained from the AVISO database on a $1/4^{\circ}\times 1/4^{\circ}$ Mercator grid 
(\url{http://www.aviso.altimetry.fr}).
The AVISO database combines altimetric data from the TOPEX/POSEIDON mission, from Jason-1 for the data after 
December, 2001 and from Envisat for the data after March, 2002. Because of sea ice coverage in the western OS, 
the altimetry data collected between January and April were out of the results and discussion. 
Oceanographic and daily wind data were provided by the World Ocean Database (WOD13) 
(\url{https://www.nodc.noaa.gov/OC5/WOD}), Pacific Oceanological Institute database 
(\url{http://oias.poi.dvo.ru}) and NCEP reanalysis (\url{http://www.esrl.noaa.gov}).
In our study we used the scatterometer-derived wind vectors and wind stress curl data 
(\url{ftp://numbat.coas.oregonstate.edu/pub/scow}) \citep{Risien2008} and MODIS SST 
satellite imagery (\url{http://oceandata.sci.gsfc.nasa.gov}).

All the Lagrangian simulation results have been obtained by solving advection equations for a 
large number of synthetic particles (tracers) advected by the AVISO velocity field
\begin{equation}
\frac{d \lambda}{d t} = u(\lambda,\varphi,t),\qquad \frac{d \varphi}{d t} =v(\lambda,\varphi,t),
\label{adveq}
\end{equation}
where $u$ and $v$ are angular zonal and meridional velocities, $\varphi$ and $\lambda$ are latitude and
longitude, respectively. Bicubical spatial interpolation and third order Lagrangian polynomials in time
are used to provide numerical results. Lagrangian trajectories are computed by integrating the equations
(\ref{adveq}) with a fourth-order Runge-Kutta scheme.

In order to track the origin of waters along the eastern SI coast in the warm and cold seasons, we have 
computed so-called Lagrangian drift maps with boundaries \citep{Prants2013,P13,Prants2014,Prants2015a}. 
A domain in the Sea is seeded at a fixed date with a large number of tracers whose trajectories are 
computed backward in time for a given period of time \citep{Prants2015b}. The waters, that entered 
a given area for that period through different geographical boundaries, are shown by different 
colors on such a map.

For computation of the meridional volume transport of the ESC, $M_y$, the Sverdrup relation
is applied
\begin{equation}
M_y = \int \beta^{-1} \rho^{-1} \operatorname{curl}_z\tau  dx,
\label{Sverdrup}
\end{equation}
where the wind stress equals to $\operatorname{curl}_z\tau = 
\partial \tau_y/\partial x - \partial \tau_x\partial y$, $\rho$ is the density of water 
and  $\beta$ is the meridional derivative of the Coriolis parameter.  
The integration path is taken along the latitudinal lines from the eastern, \E{154},   
to western,  \E{144}, boundaries of the OS. 
We used the meridional, $\tau_y$, and zonal, $\tau_x$, monthly-averaged wind stress data 
from the NCEP reanalysis. 

\section{Results}\label{results}
\begin{figure*}[!htb]
\begin{center}
\includegraphics[width=0.8\textwidth,clip]{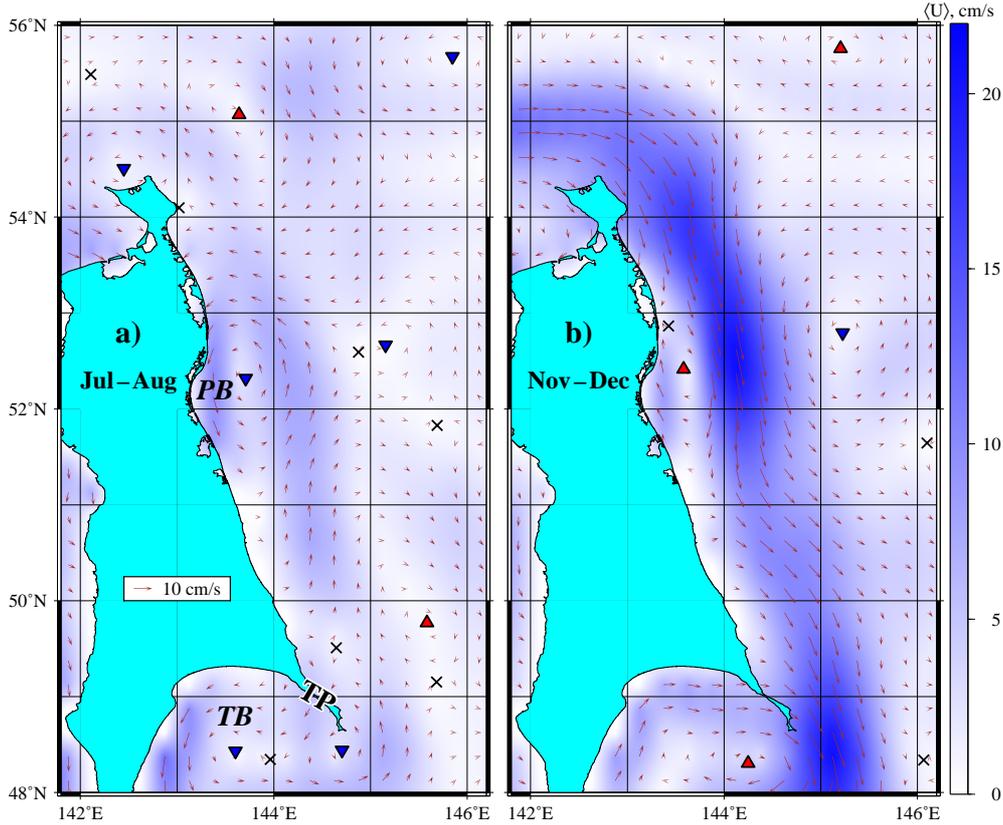}
\end{center}
\caption{The altimetric AVISO velocity field around the eastern coast of the Sakhalin Island averaged
for a) July\mdash August and b) November\mdash December from 1993 to 2016. The magnitudes 
of the averaged velocity, $\le U \ge$, are coded by nuances of the grey color. 
The centers of mesoscale cyclones and anticyclones are shown by the triangles with downward 
and upward orientation of one of the triangle's top, respectively. PB and TB stand for 
the Piltun Bay and Terpeniya Bay, respectively. TP is for the Terpeniya Peninsula.}
\label{fig2}
\end{figure*}

In July\mdash August and November\mdash December, the surface circulation along the eastern SI coast 
is determined by the mesoscale cyclones and anticyclones, respectively, located off the Piltun Bay and 
eastward of the Terpeniya Bay (Fig.~\ref{fig2}). We call them as the Piltun and Terpeniya circulation 
cells which are clearly visible in Fig.~\ref{fig2} in the altimetric AVISO velocity field averaged for 
July\mdash August (Fig.~\ref{fig2}a) and November\mdash December (Fig.~\ref{fig2}b) from 1993 to 2016. 
In the warm period, both the cells have a cyclonic circulation (Fig.~\ref{fig2}a), whereas in the cold 
period they are anticyclones with the diameter of 100~km (Fig.~\ref{fig2}b).
\begin{figure*}[!htb]
\begin{center}
\includegraphics[width=0.9\textwidth,clip]{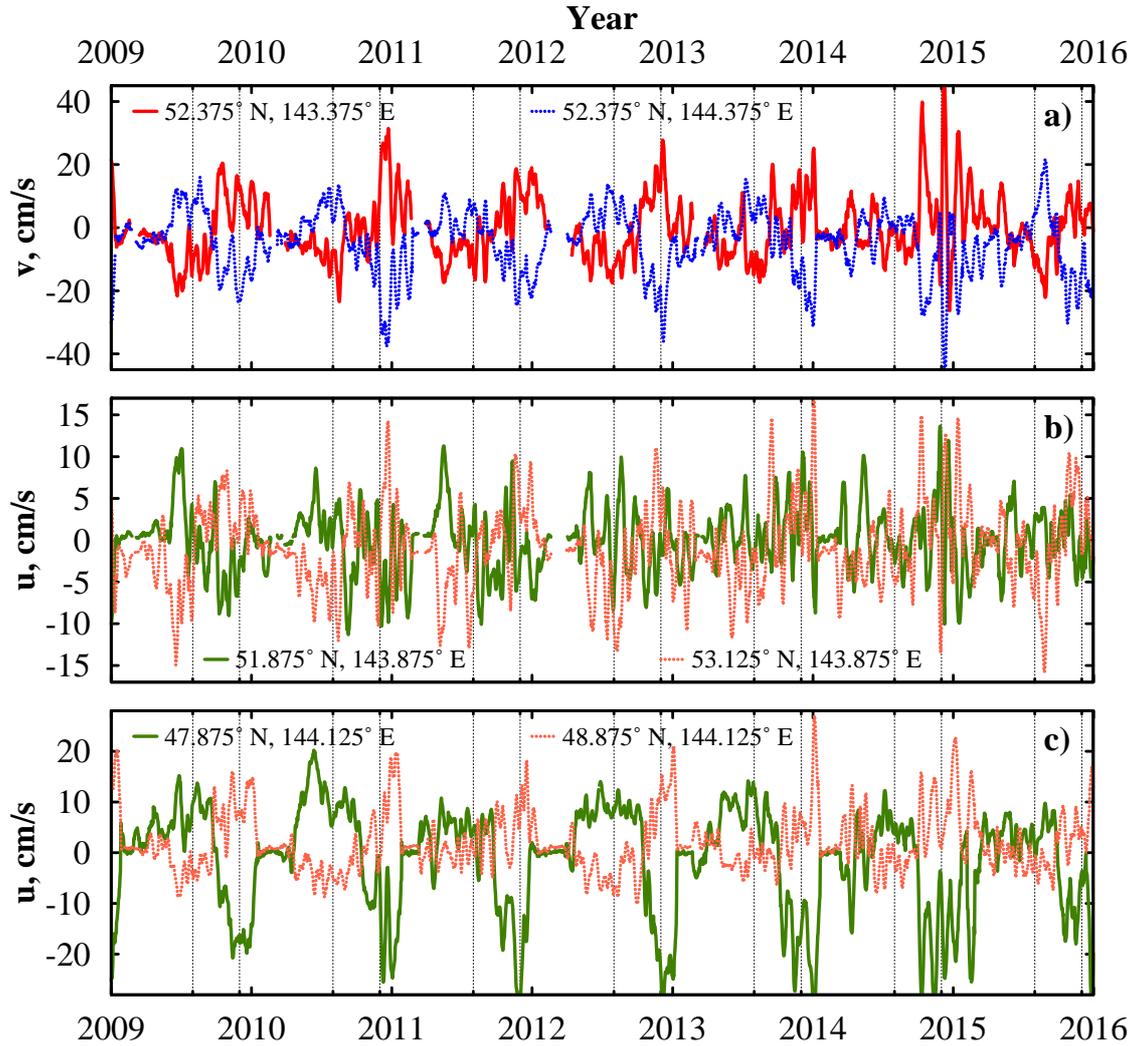}
\end{center}
\caption{a) The temporal changes in the meridional velocities at the western 
(\N{52.375}, \E{143.375}, the bold curve) and eastern (\N{52.375}, \E{144.375}, 
the dashed curve) boundaries of the Piltun circulation cell. 
b) The temporal changes in the zonal velocities at the northern (\N{53.125}, \E{143.875}, 
the dashed curve) and southern (\N{51.875}, \E{143.875}, the bold curve) 
boundaries of the Piltun circulation cell. c) The temporal changes in the zonal velocities 
at the northern (\N{48.875}, \E{144.125}, the dashed curve) and 
southern (\N{47.875}, \E{144.125}, the bold curve) boundaries of the Terpeniya circulation cell.
The vertical straight lines mark August,~1 and December,~1 for each year which are 
the middles of the warm and cold seasons in the area.}
\label{fig3}
\end{figure*}

It is useful to compute locations in the AVISO field where the velocity is zero. The standard stability
analysis allows to specify stagnation elliptic and hyperbolic points the motion around which is stable or 
unstable, respectively. The elliptic points are situated mainly at the centers of eddies. The motion around 
them is stable and circular. The hyperbolic points, situated mainly between and around eddies, are unstable 
ones with the directions along which waters converge to such a point and another directions along which they 
diverge. We mark the elliptic points on the Lagrangian maps by triangles and the hyperbolic ones~--- by crosses. 
Up (down)ward orientation of one of the triangle's top marks anticyclonic (cyclonic) rotation of water around 
them. For convenience triangles are colored in the online version as red (blue) marking centers of anticyclones 
(cyclones). The stagnation points are moving Eulerian features and may undergo bifurcations in the course of time. 
In spite of nonstationarity of the velocity field some of them may exist for weeks and much more
\citep{Prants2014d,Prants2015b}.

The AVISO velocity field, shown in Fig.~\ref{fig2}, was averaged for the warm (Fig.~\ref{fig2}a) and cold 
months (Fig.~\ref{fig2}b) for the last 23 years. So, the triangles in Fig.~\ref{fig2}a with the coordinates 
(\N{52.3}, \E{143.6}) and (\N{48.4}, \E{143.6}) specify average positions of the centers of the Piltun and 
Terpeniya cyclonic circulation cells, respectively, which regularly appear there in warm seasons. 
The triangles in Fig.~\ref{fig2}b with the coordinates (\N{52.4}, \E{143.5}) and (\N{48.3}, \E{144.2}) specify 
average positions of the centers of the Piltun and Terpeniya anticyclonic circulation cells, respectively, 
which regularly appear there in cold seasons.

\begin{figure*}[!htb]
\begin{center}
\includegraphics[width=0.85\textwidth,clip]{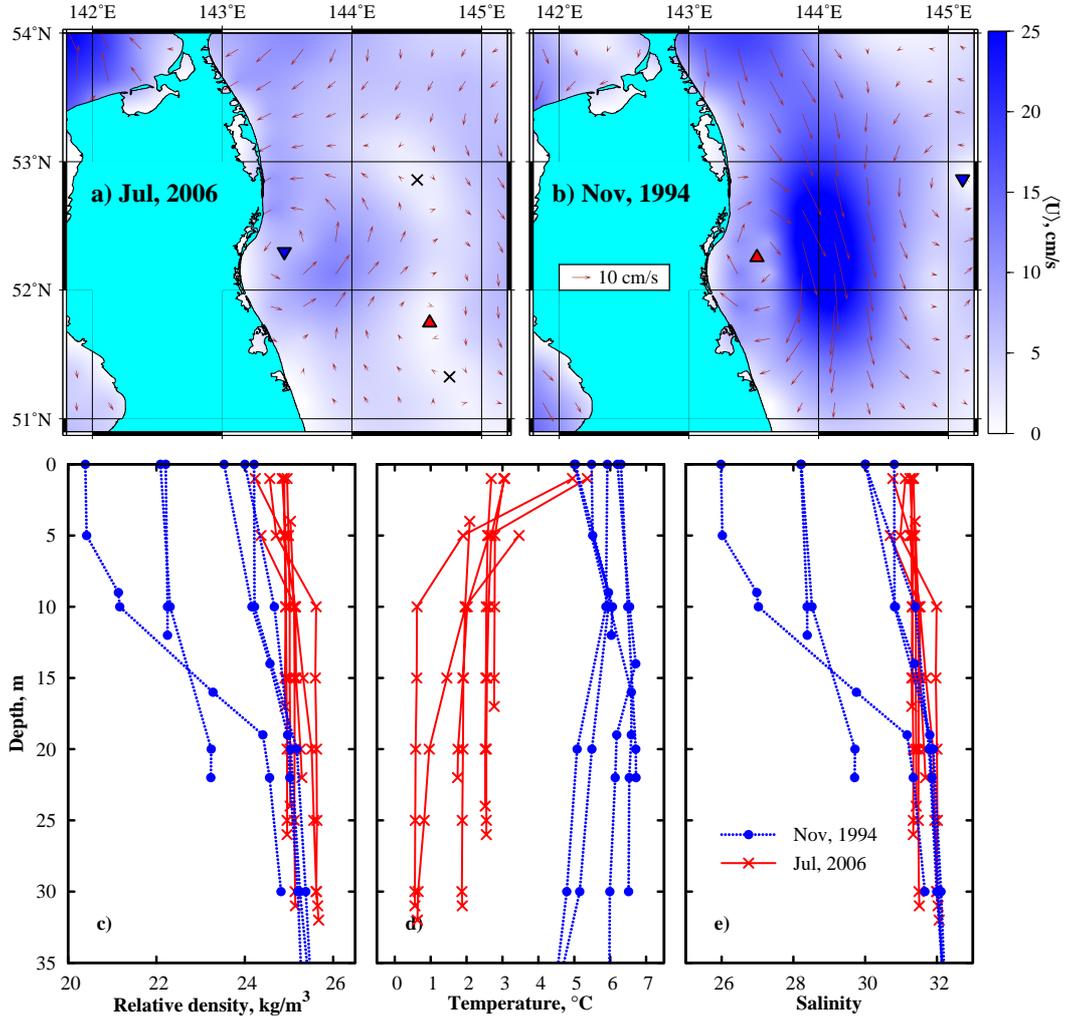}
\end{center}
\caption{The case study of the Piltun circulation cell. The altimetric AVISO velocity field 
averaged for a) July 2006 and b) November 1994 show cyclonic and anticyclonic circulations, 
respectively. The vertical 
profiles of c) the relative density, d) temperature and e) salinity obtained in the oceanographic surveys 
in July 2006 (red) and November 1994 (blue).}
\label{fig4}
\end{figure*}

In order to illustrate seasonal variations in meridional and zonal AVISO velocities at the boundaries of 
the Piltun and Terpeniya circulation cells, we plot their variations in Fig.~\ref{fig3} for the period from 2009 to 2016. 
In Fig.~\ref{fig3}a the temporal changes in the meridional velocities are shown at the western 
(\N{52.375}, \E{143.375}) and eastern (\N{52.375}, \E{144.375}) boundaries of the Piltun circulation cell. 
In Fig.~\ref{fig3}b we show the temporal changes in the zonal velocities at the northern
(\N{53.125}, \E{143.875}) and southern (\N{51.875}, \E{143.875}) boundaries of the Piltun circulation cell. The temporal 
changes in the zonal velocities at the northern (\N{48.875}, \E{144.125}) and southern (\N{47.875}, \E{144.125}) 
boundaries of the Terpeniya circulation cell are shown in Fig.~\ref{fig3}c. The vertical straight lines mark 
August,~1 and December,~1 for each year which are the middles of the warm and cold seasons in the area. 
The surface velocities are equal to 0.1--0.2 (0.2--0.3) m/s (Fig.~\ref{fig3}a and~b) at the boundaries 
of the cyclones (anticyclones). In July\mdash August (November\mdash December), the surface flow of the 
ESC is directed northward (southward) along the SI slope.

The meridional velocities at the western and eastern boundaries (Fig.~\ref{fig3}a) and zonal velocities at the 
northern and southern boundaries (Fig.~\ref{fig3}b and c) of the mesoscale circulation cells off the 
northeastern and southeastern SI undergo significant seasonal variations with its positive (negative) 
values in May\mdash September and its negative (positive) values in October\mdash December. The correlation 
coefficient between the meridional velocities at the eastern and western boundaries of the mesoscale Piltun 
circulation cell is $-0.91$ (2009--2016). The correlation coefficients between the meridional velocities at 
the eastern boundary and the zonal velocities at the northern (southern) boundaries are $-0.60$ and $0.74$. 
The seasonal changes of the flow direction at the boundaries of the mesoscale circulation cells demonstrate 
that the mesoscale cyclones are formed in the eastern SI coast area predominantly during summer, whereas 
anticyclones are generated predominantly during fall and winter.

The comparison of the geostrophical currents ($5/1000$~dbar, July 1994) off the eastern SI, computed 
with the help of the detailed CTD survey data \citep{Verkhunov1997}, with the altimetric AVISO velocity 
distributions (July~16, 1994) demonstrates a good agreement. Both the velocity fields show the existence 
of the mesoscale cyclones off the northeastern and southeastern SI (${\sim}$\N{49} and ${\sim}$\N{52}) 
and mesoscale anticyclone off the northeastern SI (${\sim}$\N{55}).

The oceanographic surveys across the Piltun circulation cell have been carried out in July 2006 and 
November 1994 when it was cyclonic (Fig.~\ref{fig4}a) and anticyclonic (Fig.~\ref{fig4}b), respectively. 
The vertical distributions of the relative density, temperature and salinity, collected across the Piltun 
cyclone in July 2006 (Fig.~\ref{fig4}c,~d and~e), show that the cyclone was composed of 
relatively low temperature, high salinity and high density waters. These waters were originated from a 
subsurface layer of the OS pelagic area. The uniform vertical distributions of the temperature and salinity 
in the cyclone core could be an indicator of the importance of tidal mixing at the SI shelf. In upper 10~m 
layer the difference in temperature and salinity between the waters, located inside and outside of the cyclone,  
exceeded 3~$^\circ$C and 2~pss, respectively. In November 1994 the anticyclone was composed of relatively 
low salinity (26--31~pss) and low density waters (20--24~kg/m$^3$). The salinity of surface (0--20~m) 
waters outside of the anticyclone (to the east) was 2--5 pss~higher than that inside it.

\begin{figure*}[!p]
\begin{center}
\includegraphics[width=0.7\textwidth,clip]{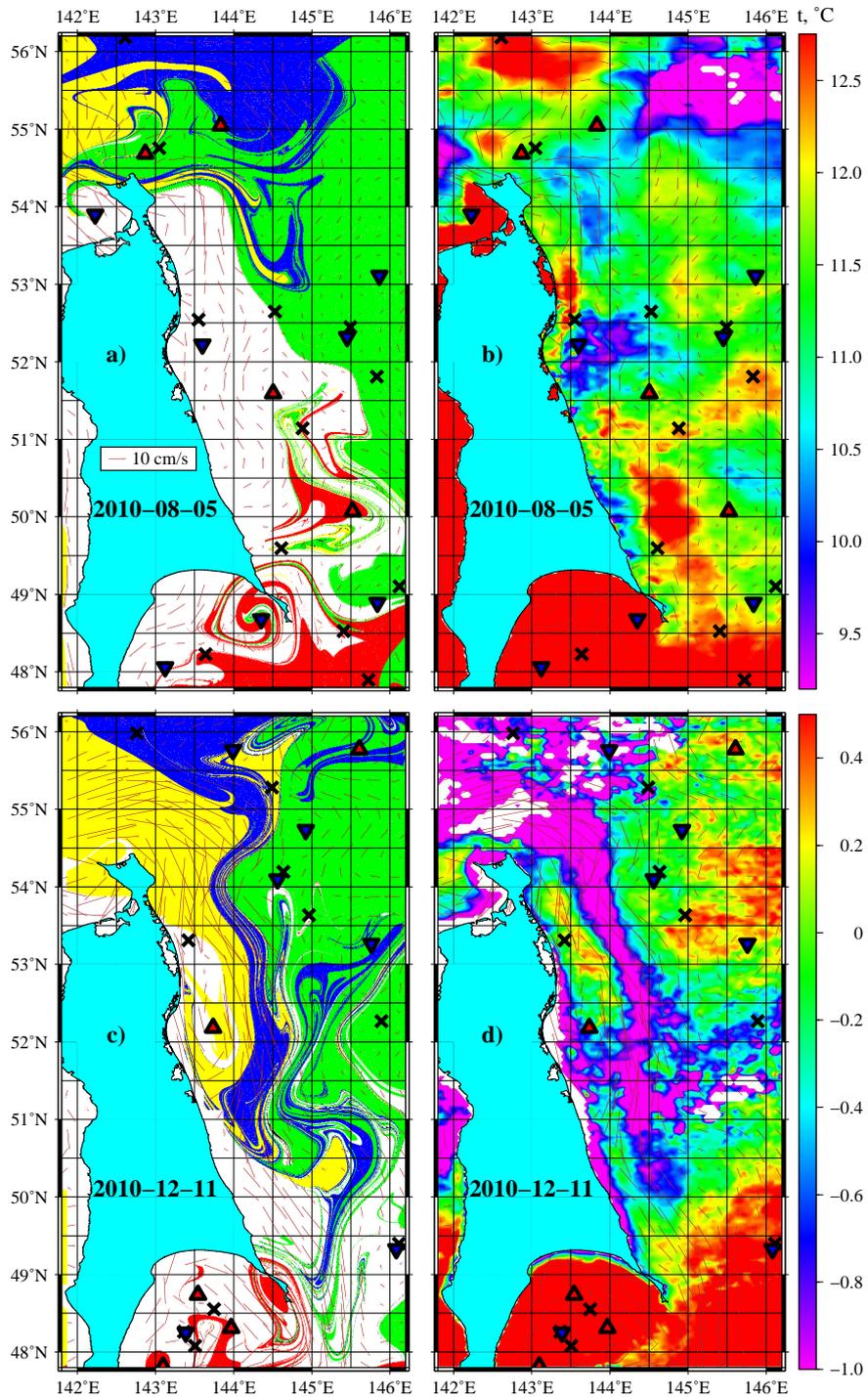}
\end{center}
\caption{a) and c) Lagrangian drift maps and b) and d) SST images (MODIS data) show examples of the 
seasonal variability of the flow along the eastern coast of the Sakhalin Island. a) and b) Formation 
of the cyclonic Piltun and Terpeniya mesoscale cells in summer. Intrusion of the warm ``red'' water 
from the southern Okhotsk Sea to the north along the Sakhalin slope is shown. c) and d) Formation of 
the anticyclonic Piltun and Terpeniya mesoscale cells in winter. Intrusion of the low-temperature 
``blue'' water from the northern Okhotsk Sea to the south along the Sakhalin slope is shown. 
SST is averaged for 3 days before and after the date indicated.}
\label{fig5}
\end{figure*}

\begin{figure*}[!htb]
\begin{center}
\includegraphics[width=0.88\textwidth,clip]{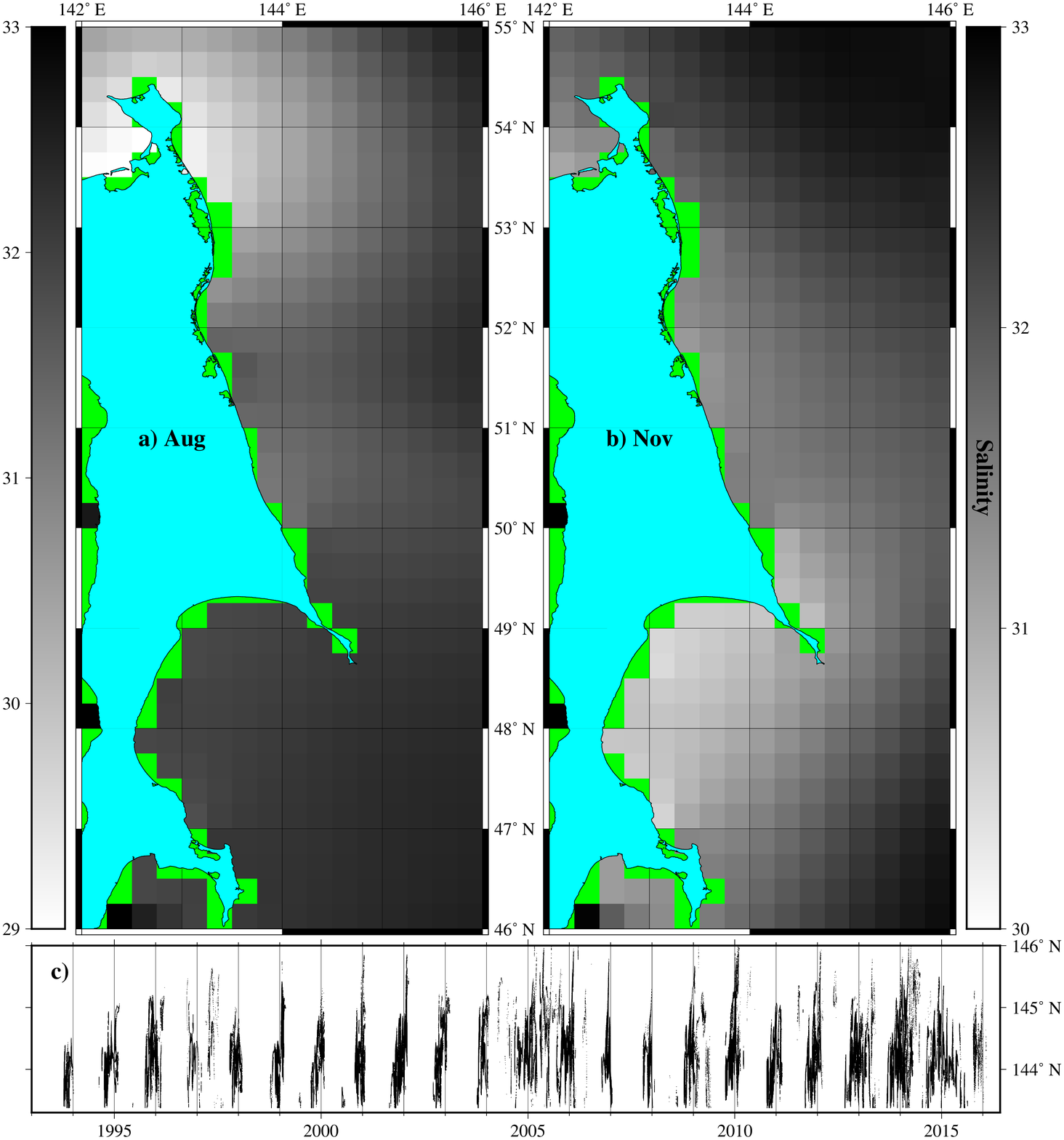}
\end{center}
\caption{Sea-surface salinity distributions averaged for a) August and b) November  
for the last 60 years (the WOD 2013 database). The values of salinity are coded by nuances 
of the grey color. Arrival of the low salinity 
waters (the light grey ones) from the Sakhalin Gulf to the the Terpeniya Bay (TB in Fig.~\ref{fig2}a) along 
the eastern Sakhalin shelf occurs in the cold season. 
c) Simulated seasonal periodicity of the southward flow along the northeastern coast of the 
Sakhalin Island. It is shown when at which longitudes the tracers launched from 1993 to 2016 
at the zonal line \N{54.25} in the Sakhalin Gulf crossed the zonal line \N{52.5} off the 
northeastern coast of the Sakhalin Island.}
\label{fig6}
\end{figure*}

The origin of waters, flowing along the eastern SI coast, can be tracked with the help of the Lagrangian 
drift maps \citep{DAN11,FAO13} which are computed backward in time as it is described in Sec.~2. The waters, that entered the
box shown in Figs.~\ref{fig5}a and~c through its western, northern, eastern 
and southern boundaries for the three months in the past, are shown by yellow, blue, green and 
red colors, respectively. 
In summer,  low salinity and warm Sakhalin Gulf waters formed by Amur River
discharge  (marked by yellow)  accumulates to the north of the SI (Fig.~\ref{fig5}a).
It is confirmed by the 
satellite SST image in Fig.~\ref{fig5}b and distribution of the mean surface salinity in Fig.~\ref{fig6}a. 
In summer, the warm ``red'' waters from the southern OS are advected to the north along the SI slope 
(Figs.~\ref{fig5}a and~b). 
%

The Lagrangian drift map  in Fig.~\ref{fig5}c, SST image in Fig.~\ref{fig5}d and mean surface salinity 
distribution in Fig.~\ref{fig6}b clearly show that in late fall less saline and relatively high temperature 
waters of the Sakhalin Gulf intrude southward from the northwestern shelf along the eastern SI coast. 
The Lagrangian map and SST imagery indicate advection of the low temperature ``blue'' waters from the 
northern OS to the south along the SI slope in winter when the anticyclonic  Piltun and Terpeniya 
mesoscale cells are formed.

The sea-surface salinity distributions, taken from the database WOD 2013, are shown in Figs.~\ref{fig6}a 
and~b with the averaging for August (panel a) and November--December (panel b) 
for the last 60 years. Arrival of the low salinity waters 
from the Sakhalin Gulf to the the Terpeniya Bay (TB in Fig.~\ref{fig2}a) along 
the eastern Sakhalin shelf occurs in the cold season. Seasonal variability 
of the flow along the eastern SI coast is illustrated in Fig.~\ref{fig6}c. We launched 100\,000 tracers each 
7 days from 1993 to 2016 along the zonal line \N{54.25} (\EE{139.7}{142.7}), crossing the Sakhalin Gulf 
(SG in Fig.~\ref{fig1}), and advected them in the AVISO velocity field. We fixed each tracer which 
crossed the zonal line \N{52.5}, the longitude of that crossing in the range from \E{142.5} to \E{146} 
and the date when it did that. The Lagrangian tracking of the Sakhalin Gulf waters clearly demonstrates 
seasonal periodicity of the southward flow along the eastern SI coast. Except for a couple of years, 
arrival of the low salinity Sakhalin Gulf waters occurred mainly in November\mdash December at the 
western SI shelf (Fig.~\ref{fig6}c).

\begin{figure*}[!htb]
\begin{center}
\includegraphics[width=0.95\textwidth,clip]{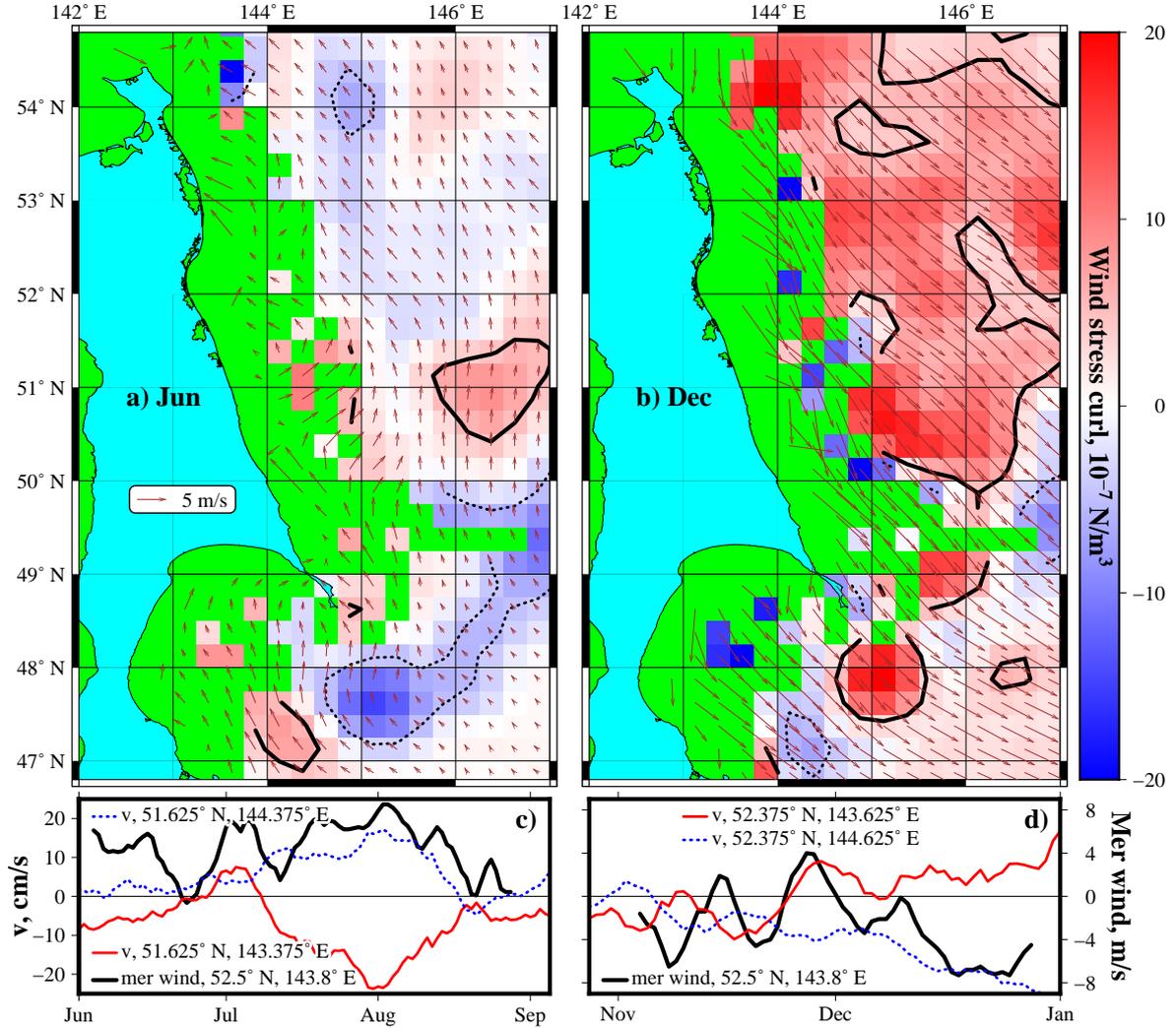}
\end{center}
\caption{Scatterometer-derived wind vectors (arrows) and wind stress curl averaged for 
a) June and b) December over the period from 1999 to 2007. Values of the wind stress 
curl are shown by the bar in the units of 10$^{-7}$ N m$^{-3}$ (colored in the online version). 
Available data are shown by the squares. Solid and dashed contour lines show the positive 
(red in the online version) and negative (blue in the online version) areas of the wind 
stress curl, respectively. Temporal changes in the AVISO meridional velocities ($v$) 
and meridional winds (daily data) in c) July\mdash August and d) November\mdash December 2013. 
AVISO meridional velocities $v$ are taken at the western (\N{51.625}, \E{143.375}) and 
eastern (\N{51.625}, \E{144.375}) boundaries of the Piltun circulation cell and shown in 
the units of cm/s by the solid and dashed curves, respectively. The velocity of 
the meridional winds is shown by the black solid curve in the units of m/s.}
\label{fig7}
\end{figure*}
\begin{figure*}[!htb]
\begin{center}
\includegraphics[width=0.8\textwidth,clip]{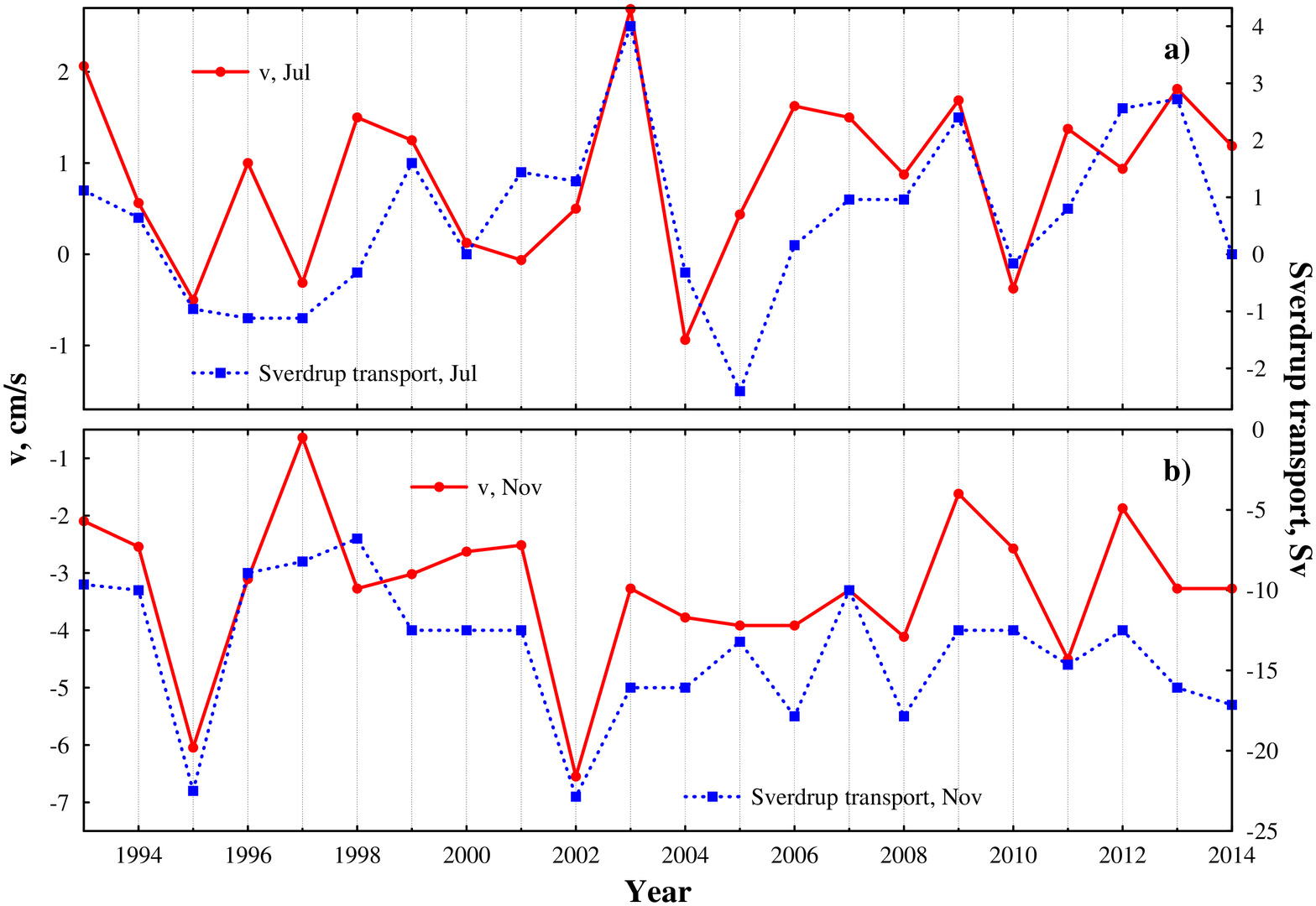}
\end{center}
\caption{Year-to-year changes of the surface meridional velocity averaged for \NN{48.38}{53.12}, 
\EE{143.38}{146.38} (solid lines) and the Sverdrup transport 
averaged for \NN{48}{54} (dotted lines).}
\label{fig8}
\end{figure*}

Prevailing winds and distribution of the wind stress curl exhibit large seasonal 
variations in the area (Figs.~\ref{fig7}a and b). In summer, the southerly (northward) winds 
along the SI are determined by the high sea level pressure center (the OS High) located over 
the OS. In November\mdash December, the Aleutian Low develops in the northern Pacific, and 
strong northerly (southward) or northwesterly winds appear along the SI coast with a positive 
wind stress curl over the northern and central OS. Due to orography and wind intensification around the capes, the northward 
winds in summer and southward winds in fall and winter lead to the local cyclonic (June) and 
anticyclonic (December) wind stress curl fields along the eastern SI coast. Figures~\ref{fig7}c 
and~d show temporal changes in the meridional wind and meridional velocities (daily data) 
at the western and eastern boundaries of the mesoscale Piltun circulation cells in 
July\mdash August and November\mdash December 2013.

The mesoscale cyclones (anticyclones) off the SI have been observed mainly during the 
periods with prevailing northward (southward) winds favorable for upwelling (downwelling). 
At the beginning of June 2013, a weak cyclonic Piltun circulation with the meridional 
velocities of 0.03--0.08~m/s has been formed due to northward winds. Northward winds 
with the velocity of about 20~m/s during July led to generation of a mesoscale cyclone 
with the meridional velocities of 0.15--0.30~m/s at its boundaries. In the third decade of 
August after the northward winds died down, intensity of the cyclonic circulation 
decreased significantly. Intensification of the winter monsoon in the middle of 
December 2013 led to formation of the anticyclonic Piltun circulation, amplification 
of the north-directed currents at the shelf and south-going component of the currents along the slope.

\section{Discussion}\label{discuss}

Figure ~\ref{fig8} shows the year-to-year changes of the surface meridional velocity 
averaged for \NN{48.38}{53.12} and \\  
\EE{143.38}{146.38} and the Sverdrup volume transport 
computed by eq.~\ref{Sverdrup} and averaged for \NN{48}{54} in July and November. 
Because of the positive wind stress curl in November, a cyclonic circulation occurs over 
the northern and central OS, and the ESC is directed southward along the eastern Sakhalin coast 
(Fig.~\ref{fig8}b). 
Its surface velocity is determined by values of the wind stress curl with the correlation coefficient 
$r= 0.78$. 
Figure~\ref{fig8}a demonstrates that the direction and magnitude of the ESC surface 
velocity in July is determined by the wind stress curl over the OS with $r= 0.60$.  
When wind stress curl over the OS is negative (anticyclonic) or positive (cyclonic) in July, 
ESC surface velocity tends to be northward or southward, respectively. We may assume that that spin-up of 
the cyclonic gyre in the OS, forced by a positive wind stress curl over the northern part of 
the Sea in fall and winter, intensifies the southward transport 
of the ECS and causes a mesoscale anticyclonic circulation off the northeastern SI. 

The wind stress curl fields from QuikSCAT show that local anticyclonic (and cyclonic) wind 
stress maxima along the SI coast (Fig.~\ref{fig7}) are associated with the anticyclonic 
(and cyclonic) circulation cells (Figs.~\ref{fig2} and \ref{fig3}). Our results are consistent 
with the hypothesis of anticyclonic eddy formation due to local anticyclonic wind stress curl 
associated with orography and wind intensification around the capes \citep{Perlin2004}.

In summer, the low salinity waters, formed by Amur-River discharge, are concentrated in the 
coastal area of the northern SI (Figs.~\ref{fig5} and~\ref{fig6}). In fall and winter, during 
winter monsoon, the low salinity and low density waters are transporting by the ESC from the 
northern end of the SI to the southern part of the OS along the eastern SI coast. Therefore, 
both barotropic and baroclinic effects could contribute to intensification of the southward 
flow of the ESC and mesoscale anticyclone generation in fall\mdash winter. Correlation 
between the meridional velocities at the boundaries of the mesoscale cyclone and the meridional 
wind in summer (Figs.~\ref{fig7}a and~c) could be related to the strength of the upwelling events 
that occur along the shelf edge of the eastern SI. The wind, directed northward along the eastern 
SI coast, generates an Ekman transport of water in the eastern direction toward the deep basin 
and hence causes an upwelling of abyssal waters \citep{Verkhunov1997, Rutenko2009, Rutenko2014}.
The alongshore upwelling fronts have been often observed in satellite infrared images of the OS 
in summer when the presence of the seasonal thermocline provides pronounced contrasts of the SST 
(Fig.~\ref{fig5}b). The upwelling in the northern and central parts of the eastern SI shelf took 
place from July to August. In the southern part of the region (off the Terpeniya Peninsula),  
upwelling occurs during August\mdash September. Variability of the upwelling is primarily 
driven by regional wind forcing \citep{Zhabin2016}. Decrease of the sea level and shallowing 
the density contours toward the shore (westward), forced by the Ekman drift, lead to the 
northward flow of the ESC in summer \citep{Verkhunov1997}. The upwelling often results in 
formation of cyclonic eddies. The eddies can be formed by baroclinic 
instability of the alongshore jet (generated by the upwelling) regardless to the existence 
of topographic irregularities \citep{Zhurbas2006}.

The upwelling enriches the euphotic layer of the SI shelf (around 50~m) with nutrients and thereby 
could promote the phytoplankton bloom. Our results indicate that the extremely high biological 
productivity (4--6~g~C m$^{-2}$~day$^{-1}$) and high concentration of chlorophyll-a (3--7~$\mu$g/l),  
observed in July\mdash August 1993, 1994 and 2003 \citep{Sorokin_2002, Belan2005}, are related to 
the mesoscale cyclones at the shelf edge off the northeastern SI (Piltun cyclones). The mesoscale 
cyclones tend to upwell nutrient-enriched deep waters into the euphotic zone thereby increasing 
biological community production \citep{Falkowski1991}. The extremely high concentration of 
chlorophyll-a, observed in the mesoscale cyclones off the northeastern SI in July\mdash August, 
has been probably caused by high nutrient concentrations (NO$_3$ ${\sim}$10--20~$\mu$mol/kg, 
SiO$_2$ ${\sim}$15--40~$\mu$mol/kg) and low temperatures (from $-2$ to 1~$^\circ$C) providing  
comfortable conditions for the large diatom phytoplankton growth.

\section{Conclusions}\label{concl}
The flow field in the ESC area is characterized by existence of the mesoscale circulation cells. 
Formation of the anticyclonic (November\mdash December) and cyclonic (July\mdash August) mesoscale 
circulations in this  area causes seasonal flow reversals along the East Sakhalin shelf and slope. 
The strong seasonality in surface circulation could be explained by temporal changes in the wind 
stress curl and wind direction along the Sakhalin. The mesoscale cyclones generation is related 
to a coastal upwelling forcing by northward winds and positive wind stress curl along the Sakhalin 
coast. The anticyclones formation is related to inflow of low salinity waters from the Sakhalin Gulf 
forced by southward winds and negative wind stress curl along the Sakhalin coast. The mesoscale 
cyclones and anticyclones provide water exchange between the shelf and deep basin of the OS.
The mesoscale cyclonic circulation can been considered as direct and indirect causes of an 
increase of the biological productivity at the northeast Sakhalin shelf in summer.

\section*{Acknowledgments}
This work was supported by the Russian Science Foundation (project no.~16--17--10025). 
The altimeter products were distributed by AVISO with support from CNES.

\bibliographystyle{model2-names}      
\bibliography{elsarticle-template-2-harv}{}

\begin{thebibliography}{25}
\expandafter\ifx\csname natexlab\endcsname\relax\def\natexlab#1{#1}\fi
\providecommand{\url}[1]{\texttt{#1}}
\providecommand{\href}[2]{#2}
\providecommand{\path}[1]{#1}
\providecommand{\DOIprefix}{doi:}
\providecommand{\ArXivprefix}{arXiv:}
\providecommand{\URLprefix}{URL: }
\providecommand{\Pubmedprefix}{pmid:}
\providecommand{\doi}[1]{\href{http://dx.doi.org/#1}{\path{#1}}}
\providecommand{\Pubmed}[1]{\href{pmid:#1}{\path{#1}}}
\providecommand{\bibinfo}[2]{#2}
\ifx\xfnm\relax \def\xfnm[#1]{\unskip,\space#1}\fi
\bibitem[{Belan et~al.(2005)Belan, Budaeva, Makarov, Propp, Selina, Orlova and
  Stonik}]{Belan2005}
\bibinfo{author}{Belan, T.A.}, \bibinfo{author}{Budaeva, V.D.},
  \bibinfo{author}{Makarov, V.G.}, \bibinfo{author}{Propp, L.N.},
  \bibinfo{author}{Selina, M.S.}, \bibinfo{author}{Orlova, T.Y.},
  \bibinfo{author}{Stonik, I.V.}, \bibinfo{year}{2005}.
\newblock \bibinfo{title}{Oceanographical and hydrobiological investigations
  along north east {S}akhalin {I}sland in summer 2003}.
\newblock \bibinfo{journal}{Pacific Oceanography} \bibinfo{volume}{3},
  \bibinfo{pages}{66--69}.
\bibitem[{Ebuchi(2006)}]{Ebuchi2006}
\bibinfo{author}{Ebuchi, N.}, \bibinfo{year}{2006}.
\newblock \bibinfo{title}{Seasonal and interannual variations in the {E}ast
  {S}akhalin current revealed by {TOPEX}/{POSEIDON} altimeter data}.
\newblock \bibinfo{journal}{Journal of Oceanography} \bibinfo{volume}{62},
  \bibinfo{pages}{171--183}.
\newblock \DOIprefix\doi{10.1007/s10872-006-0042-x}.
\bibitem[{Falkowski et~al.(1991)Falkowski, Ziemann, Kolber and
  Bienfang}]{Falkowski1991}
\bibinfo{author}{Falkowski, P.G.}, \bibinfo{author}{Ziemann, D.},
  \bibinfo{author}{Kolber, Z.}, \bibinfo{author}{Bienfang, P.K.},
  \bibinfo{year}{1991}.
\newblock \bibinfo{title}{Role of eddy pumping in enhancing primary production
  in the ocean}.
\newblock \bibinfo{journal}{Nature} \bibinfo{volume}{352},
  \bibinfo{pages}{55--58}.
\newblock \DOIprefix\doi{10.1038/352055a0}.
\bibitem[{Kochergin et~al.(1999)Kochergin, Rybalko, Putov and
  Shevchenko}]{Kochergin1999}
\bibinfo{author}{Kochergin, I.E.}, \bibinfo{author}{Rybalko, S.I.},
  \bibinfo{author}{Putov, V.F.}, \bibinfo{author}{Shevchenko, G.V.},
  \bibinfo{year}{1999}.
\newblock \bibinfo{title}{Hydrometeorological and ecological conditions of the
  {F}ar-{E}astern {S}eas: marine environmental impact assessment. FERHRI
  special issue 2}. \bibinfo{publisher}{Dalnauka},
  \bibinfo{address}{Vladivostok}. Chapter \bibinfo{chapter}{Processing of the
  instrumental current data collected in the {P}iltun-{A}stokh and
  {A}rkutun-{D}agi oil fields}.
\newblock pp. \bibinfo{pages}{96--113}.
\newblock \bibinfo{note}{[in Russian]}.
\bibitem[{Kusailo et~al.(2013)Kusailo, Shevchenko and Chastikov}]{Kusailo2013}
\bibinfo{author}{Kusailo, O.V.}, \bibinfo{author}{Shevchenko, G.V.},
  \bibinfo{author}{Chastikov, V.N.}, \bibinfo{year}{2013}.
\newblock \bibinfo{title}{Extreme nonperiodic currents on the northeastern
  shelf of {S}akhalin island}.
\newblock \bibinfo{journal}{Doklady Earth Sciences} \bibinfo{volume}{448},
  \bibinfo{pages}{97--102}.
\newblock \DOIprefix\doi{10.1134/s1028334x1301011x}.
\bibitem[{Meier et~al.(2007)Meier, Yazvenko, Blokhin, Wainwright, Maminov,
  Yakovlev and Newcomer}]{Meier2007}
\bibinfo{author}{Meier, S.K.}, \bibinfo{author}{Yazvenko, S.B.},
  \bibinfo{author}{Blokhin, S.A.}, \bibinfo{author}{Wainwright, P.},
  \bibinfo{author}{Maminov, M.K.}, \bibinfo{author}{Yakovlev, Y.M.},
  \bibinfo{author}{Newcomer, M.W.}, \bibinfo{year}{2007}.
\newblock \bibinfo{title}{Distribution and abundance of western gray whales off
  northeastern {S}akhalin {I}sland, {R}ussia, 2001--2003}.
\newblock \bibinfo{journal}{Environmental Monitoring and Assessment}
  \bibinfo{volume}{134}, \bibinfo{pages}{107--136}.
\newblock \DOIprefix\doi{10.1007/s10661-007-9811-2}.
\bibitem[{Moroshkin(1966)}]{Moroshkin1966}
\bibinfo{author}{Moroshkin, K.V.}, \bibinfo{year}{1966}.
\newblock \bibinfo{title}{Water masses of the {O}khotsk {S}ea}.
\newblock \bibinfo{publisher}{Nauka}, \bibinfo{address}{Moscow}.
\newblock \bibinfo{note}{[in Russian]}.
\bibitem[{Ohshima et~al.(2004)Ohshima, Simizu, Itoh, Mizuta, Fukamachi, Riser
  and Wakatsuchi}]{Ohshima2004}
\bibinfo{author}{Ohshima, K.I.}, \bibinfo{author}{Simizu, D.},
  \bibinfo{author}{Itoh, M.}, \bibinfo{author}{Mizuta, G.},
  \bibinfo{author}{Fukamachi, Y.}, \bibinfo{author}{Riser, S.C.},
  \bibinfo{author}{Wakatsuchi, M.}, \bibinfo{year}{2004}.
\newblock \bibinfo{title}{Sverdrup balance and the cyclonic gyre in the {S}ea
  of {O}khotsk}.
\newblock \bibinfo{journal}{Journal of Physical Oceanography}
  \bibinfo{volume}{34}, \bibinfo{pages}{513--525}.
\newblock \DOIprefix\doi{10.1175/1520-0485(2004)034<0513:sbatcg>2.0.co;2}.
\bibitem[{Perlin et~al.(2004)Perlin, Samelson and Chelton}]{Perlin2004}
\bibinfo{author}{Perlin, N.}, \bibinfo{author}{Samelson, R.M.},
  \bibinfo{author}{Chelton, D.B.}, \bibinfo{year}{2004}.
\newblock \bibinfo{title}{Scatterometer and model wind and wind stress in the
  {O}regon\,--\,{N}orthern {C}alifornia coastal zone}.
\newblock \bibinfo{journal}{Monthly Weather Review} \bibinfo{volume}{132},
  \bibinfo{pages}{2110--2129}.
\newblock \DOIprefix\doi{10.1175/1520-0493(2004)132<2110:samwaw>2.0.co;2}.
\bibitem[{Pishchalnik and Arhipkin(1999)}]{Pishchalnik1999}
\bibinfo{author}{Pishchalnik, V.M.}, \bibinfo{author}{Arhipkin, V.S.},
  \bibinfo{year}{1999}.
\newblock \bibinfo{title}{Hydrometeorological and ecological conditions of the
  {F}ar-{E}astern {S}eas: marine environmental impact assessment. FERHRI
  special issue 2}. \bibinfo{publisher}{Dalnauka},
  \bibinfo{address}{Vladivostok}. Chapter \bibinfo{chapter}{Seasonal variations
  of the circulation on {S}akhalin shelf}.
\newblock pp. \bibinfo{pages}{84--95}.
\newblock \bibinfo{note}{[in Russian]}.
\bibitem[{Prants(2013)}]{P13}
\bibinfo{author}{Prants, S.V.}, \bibinfo{year}{2013}.
\newblock \bibinfo{title}{Dynamical systems theory methods to study mixing and
  transport in the ocean}.
\newblock \bibinfo{journal}{Physica Scripta} \bibinfo{volume}{87},
  \bibinfo{pages}{038115}.
\newblock \DOIprefix\doi{10.1088/0031-8949}.
\bibitem[{Prants(2014)}]{Prants2014d}
\bibinfo{author}{Prants, S.V.}, \bibinfo{year}{2014}.
\newblock \bibinfo{title}{Chaotic {L}agrangian transport and mixing in the
  ocean}.
\newblock \bibinfo{journal}{The European Physical Journal Special Topics}
  \bibinfo{volume}{223}, \bibinfo{pages}{2723--2743}.
\newblock \DOIprefix\doi{10.1140/epjst/e2014-02288-5}.
\bibitem[{Prants(2015)}]{Prants2015b}
\bibinfo{author}{Prants, S.V.}, \bibinfo{year}{2015}.
\newblock \bibinfo{title}{Backward-in-time methods to simulate large-scale
  transport and mixing in the ocean}.
\newblock \bibinfo{journal}{Physica Scripta} \bibinfo{volume}{90},
  \bibinfo{pages}{074054}.
\newblock \DOIprefix\doi{10.1088/0031-8949/90/7/074054}.
\bibitem[{Prants et~al.(2013a)Prants, Andreev, Budyansky and
  Uleysky}]{Prants2013}
\bibinfo{author}{Prants, S.V.}, \bibinfo{author}{Andreev, A.G.},
  \bibinfo{author}{Budyansky, M.V.}, \bibinfo{author}{Uleysky, M.Y.},
  \bibinfo{year}{2013}a.
\newblock \bibinfo{title}{Impact of mesoscale eddies on surface flow between
  the {P}acific {O}cean and the {B}ering {S}ea across the {N}ear {S}trait}.
\newblock \bibinfo{journal}{Ocean Modelling} \bibinfo{volume}{72},
  \bibinfo{pages}{143--152}.
\newblock \DOIprefix\doi{10.1016/j.ocemod.2013.09.003}.
\bibitem[{Prants et~al.(2015)Prants, Andreev, Budyansky and
  Uleysky}]{Prants2015a}
\bibinfo{author}{Prants, S.V.}, \bibinfo{author}{Andreev, A.G.},
  \bibinfo{author}{Budyansky, M.V.}, \bibinfo{author}{Uleysky, M.Y.},
  \bibinfo{year}{2015}.
\newblock \bibinfo{title}{Impact of the {A}laskan {S}tream flow on surface
  water dynamics, temperature, ice extent, plankton biomass, and walleye
  pollock stocks in the eastern {O}khotsk {S}ea}.
\newblock \bibinfo{journal}{Journal of Marine Systems} \bibinfo{volume}{151},
  \bibinfo{pages}{47--56}.
\newblock \DOIprefix\doi{10.1016/j.jmarsys.2015.07.001}.
\bibitem[{Prants et~al.(2014)Prants, Budyansky and Uleysky}]{Prants2014}
\bibinfo{author}{Prants, S.V.}, \bibinfo{author}{Budyansky, M.V.},
  \bibinfo{author}{Uleysky, M.Y.}, \bibinfo{year}{2014}.
\newblock \bibinfo{title}{Lagrangian study of surface transport in the
  {K}uroshio {E}xtension area based on simulation of propagation of
  {F}ukushima-derived radionuclides}.
\newblock \bibinfo{journal}{Nonlinear Processes in Geophysics}
  \bibinfo{volume}{21}, \bibinfo{pages}{279--289}.
\newblock \DOIprefix\doi{10.5194/npg-21-279-2014}.
\bibitem[{Prants et~al.(2013b)Prants, Ponomarev, Budyansky, Uleysky and
  Fayman}]{FAO13}
\bibinfo{author}{Prants, S.V.}, \bibinfo{author}{Ponomarev, V.I.},
  \bibinfo{author}{Budyansky, M.V.}, \bibinfo{author}{Uleysky, M.Y.},
  \bibinfo{author}{Fayman, P.A.}, \bibinfo{year}{2013}b.
\newblock \bibinfo{title}{Lagrangian analysis of mixing and transport of water
  masses in the marine bays}.
\newblock \bibinfo{journal}{Izvestiya, Atmospheric and Oceanic Physics}
  \bibinfo{volume}{49}, \bibinfo{pages}{82--96}.
\newblock \DOIprefix\doi{10.1134/S0001433813010088}.
\bibitem[{Prants et~al.(2011)Prants, Uleysky and Budyansky}]{DAN11}
\bibinfo{author}{Prants, S.V.}, \bibinfo{author}{Uleysky, M.Y.},
  \bibinfo{author}{Budyansky, M.V.}, \bibinfo{year}{2011}.
\newblock \bibinfo{title}{Numerical simulation of propagation of radioactive
  pollution in the ocean from the {F}ukushima {D}ai-ichi nuclear power plant}.
\newblock \bibinfo{journal}{Doklady Earth Sciences} \bibinfo{volume}{439},
  \bibinfo{pages}{1179--1182}.
\newblock \DOIprefix\doi{10.1134/S1028334X11080277}.
\bibitem[{Risien and Chelton(2008)}]{Risien2008}
\bibinfo{author}{Risien, C.M.}, \bibinfo{author}{Chelton, D.B.},
  \bibinfo{year}{2008}.
\newblock \bibinfo{title}{A global climatology of surface wind and wind stress
  fields from eight years of {QuikSCAT} scatterometer data}.
\newblock \bibinfo{journal}{Journal of Physical Oceanography}
  \bibinfo{volume}{38}, \bibinfo{pages}{2379--2413}.
\newblock \DOIprefix\doi{10.1175/2008jpo3881.1}.
\bibitem[{Rutenko et~al.(2009)Rutenko, Khrapchenkov and Sosnin}]{Rutenko2009}
\bibinfo{author}{Rutenko, A.N.}, \bibinfo{author}{Khrapchenkov, F.F.},
  \bibinfo{author}{Sosnin, V.A.}, \bibinfo{year}{2009}.
\newblock \bibinfo{title}{Near-shore upwelling on the {S}akhalin shelf}.
\newblock \bibinfo{journal}{Russian Meteorology and Hydrology}
  \bibinfo{volume}{34}, \bibinfo{pages}{93--99}.
\newblock \DOIprefix\doi{10.3103/s1068373909020058}.
\bibitem[{Rutenko and Sosnin(2014)}]{Rutenko2014}
\bibinfo{author}{Rutenko, A.N.}, \bibinfo{author}{Sosnin, V.A.},
  \bibinfo{year}{2014}.
\newblock \bibinfo{title}{Hydrodynamic processes on the {S}akhalin shelf in the
  coastal {P}iltun area of the grey whale feeding and their correlation with
  atmospheric circulation}.
\newblock \bibinfo{journal}{Russian Meteorology and Hydrology}
  \bibinfo{volume}{39}, \bibinfo{pages}{335--349}.
\newblock \DOIprefix\doi{10.3103/s1068373914050070}.
\bibitem[{Sorokin and Sorokin(2002)}]{Sorokin_2002}
\bibinfo{author}{Sorokin, Y.I.}, \bibinfo{author}{Sorokin, P.Y.},
  \bibinfo{year}{2002}.
\newblock \bibinfo{title}{Microplankton and primary production in the {S}ea of
  {O}khotsk in summer 1994}.
\newblock \bibinfo{journal}{Journal of Plankton Research} \bibinfo{volume}{24},
  \bibinfo{pages}{453--470}.
\newblock \DOIprefix\doi{10.1093/plankt/24.5.453}.
\bibitem[{Verkhunov(1997)}]{Verkhunov1997}
\bibinfo{author}{Verkhunov, A.V.}, \bibinfo{year}{1997}.
\newblock \bibinfo{title}{Complex studies of ecosystem of the {S}ea of
  {O}khotsk}. \bibinfo{publisher}{VNIRO}, \bibinfo{address}{Moscow}. Chapter
  \bibinfo{chapter}{Improvement of our knowledge about the large-scale
  circulation in the {O}khotsk {S}ea}.
\newblock Ecology of the seas of Russia, pp. \bibinfo{pages}{6--17}.
\newblock \bibinfo{note}{[in Russian]}.
\bibitem[{Zhabin and Dmitrieva(2016)}]{Zhabin2016}
\bibinfo{author}{Zhabin, I.A.}, \bibinfo{author}{Dmitrieva, E.V.},
  \bibinfo{year}{2016}.
\newblock \bibinfo{title}{Seasonal and interannual variability of wind-driven
  upwelling along eastern {S}akhalin {I}sland coast based on the
  {QuikSCAT}/{SeaWinds} scatterometer data}.
\newblock \bibinfo{journal}{Earth research from space} \bibinfo{volume}{2016},
  \bibinfo{pages}{105--115}.
\newblock \DOIprefix\doi{10.7868/s0205961416010152}. \bibinfo{note}{[in
  Russian]}.
\bibitem[{Zhurbas et~al.(2006)Zhurbas, Oh and Park}]{Zhurbas2006}
\bibinfo{author}{Zhurbas, V.}, \bibinfo{author}{Oh, I.S.},
  \bibinfo{author}{Park, T.}, \bibinfo{year}{2006}.
\newblock \bibinfo{title}{Formation and decay of a longshore baroclinic jet
  associated with transient coastal upwelling and downwelling: {A} numerical
  study with applications to the {B}altic {S}ea}.
\newblock \bibinfo{journal}{Journal of Geophysical Research: Oceans}
  \bibinfo{volume}{111}, \bibinfo{pages}{C04014}.
\newblock \DOIprefix\doi{10.1029/2005jc003079}.

\end{thebibliography}
\end{document}